# Nano-artifact metrics based on random collapse of resist

Tsutomu Matsumoto[1,a], Morihisa Hoga[2], Yasuyuki Ohyagi[2], Mikio Ishikawa[2],

Makoto Naruse[3], Kenta Hanaki[1], Ryosuke Suzuki[1], Daiki Sekiguchi[1],

Naoya Tate[4] & Motoichi Ohtsu[4]

[1]*Graduate School of Environment and Information Sciences, Yokohama National University, Hodogaya, Yokohama, Kanagawa 240-8501, Japan*

[2]*Dai Nippon Printing Co. Ltd., 250-1 Wakashiba, Kashiwa, Chiba 277-0871, Japan*

[3]*Photonic Network Research Institute, National Institute of Information and Communications Technology, 4-2-1 Nukui-kita, Koganei, Tokyo 184-8795, Japan*

[4]*Department of Electrical Engineering and Information Systems, Graduate School of Engineering, The University of Tokyo, 2-11-16 Yayoi, Bunkyo-ku, Tokyo 113-8656, Japan*

a) Electronic mail: tsutomu@ynu.ac.jp

**ABSTRACT: Artifact metrics is an information security technology that uses the intrinsic characteristics of a physical object for authentication and clone resistance. Here, we demonstrate nano-artifact metrics based on silicon nanostructures formed via an array of resist pillars that randomly collapse when exposed to electron-beam lithography. The proposed technique uses conventional and scalable lithography processes, and because of the random collapse of resist, the resultant structure has extremely fine-scale morphology with a minimum dimension below 10 nm, which is less than the resolution of current lithography capabilities. By evaluating false match, false non-match and clone-resistance rates, we clarify that the nanostructured patterns based on resist collapse satisfy the requirements for high-performance security applications.**

Artifact metrics[1] uses physical features unique to individual objects in terms of their physical properties or combinations of these properties, including electromagnetic[2,3], mechanical and optical properties[4,5]. For an artifact metric to function, it should satisfy several conditions, such as (1) the extracted characters should vary between individual objects (individuality), (2) a given response should be consistently obtained for each measurement (measurement stability), (3) they should be robust against degradation caused by common use (durability) and (4) fabricated clones having an equivalent physical characteristic should be extremely difficult (clone resistance). Examples of existing artifact metrics include ordinary paper[5], paper containing magnetic microfibers[6], plastics and semiconductor chips. A physical unclonable function[7] is a type of artifact metrics that is essentially equivalent to what Matsumoto *et al.* examined under the name of 'clone-resistant modules' in 1997[8].

The critical-security battlefield in which artifact metrics are used is analogous to a defender and attacker relationship in which the former tries to produce patterns that are difficult copies, and the latter seeks to counterfeit these patters. In view of recent technological advancements in microfabrication and its strong demand in society and industry (e.g. optical document security[9]), new technology must go beyond that developed so far, which has been limited to micrometre-scale precision, and be founded on the ultimate principles of physics. Here, we propose and demonstrate a nano-artifact metrics that is robust against cloning attacks. The proposed metric uses nanometre-scale structures obtained from the random collapse of resists induced by exposure to conventional electron-beam (e-beam) lithography.

E-beam lithography is a mature and fundamental technology for prototyping fine structures. Minimum feature size is an important metric for lithography to produce the designated structures. To quantify the achievable minimum size, we use a two-dimensional array of pillars. The decrease in the minimum pitch of a pillar array over the last few years is summarized in Fig. 1a. The circles denoted (1), (2) and (3) in Fig. 1a are based on Refs. [10], [11] and [12], respectively and the dotted line, which represents an estimated pitch-resolution limit, is based on Refs. [13] and [14]. Figure 1a also suggests that these feature sizes may be fabricated by attackers who use the available technology to make clones.

Meanwhile, we must also consider the extent to which we can precisely measure fine structures with the available technology such as a scanning electron microscopy (SEM). Critical-dimension scanning electron microscopy (CD-SEM), which is specialized in measuring length and offers precision in the sub-nanometre scale, may be assumed[15]. For an artifact metrics to be made using silicon nanostructures fabricated based on conventional e-beam lithography, the defender, who wants to prevent counterfeiting, must fabricate fine-structured patterns such that the attacker, who wants to copy the authentic device, will not be able to intentionally reproduce the pattern based on the information obtained by CD-SEM. However, this condition imposes a paradoxical requirement that fine structures which are smaller than the resolution limit of the state-of-the-art e-beam lithography should be fabricated. Otherwise, the authentic devices may be easily cloned.

To overcome this paradox, we exploit the well-known phenomenon of the random collapse of resist[16]. Resist collapse may occur during the rinse process of lithography and depends on the pattern resolution, resist thickness and duration of e-beam exposure. The end result is the collapse of the intended pattern[16]. To produce a desired pattern, resist collapse must be suppressed in e-beam lithography, which can be achieved by deployment of 'anticollapse rinses'. However, from the standpoint of nano-artifact metrics, resist collapse occasionally provides structures finer than the original technological limitation. Furthermore, resist collapse occurs randomly. Therefore, as shown in Fig. 1a, we can use resist collapse to benefit from the uncertainty in position that is less than the resolution of nanofabrication and achieve nano-artifact-metric functionalities.

To verify this notion, we fabricated an array of pillars from a layer of resist. The pillars had cross section area of 60 nm × 60 nm, were 200 nm high and were positioned on a grid of 120 nm × 120 nm squares that filled a 2 μm × 2 μm square, as shown in Fig. 1b. As a guide for facilitating alignment, a 3 μm × 3 μm square frame was drawn outside the pillar array area. We used a JEOL JBX-9300FS e-beam lithography system with the acceleration voltage set at 100 kV and with a dose of 37 μC/cm$^2$. After post-exposure bake and resist development, the structure is rinsed, which is when the random collapse of resist pillars occurs. Figure 1c shows an SEM image of an array of collapsed resist pillars. The wafer is then etched with HBr-based gas using inductively coupled plasma (ICP)-type reactive ion etching (RIE), and the resist is stripped by oxygen ashing. The resulting nanostructured-silicon patterns were imaged by a CD-SEM (Hitachi High-Technologies CG4000). We fabricated 2401 samples on a single 200-mm-diamter wafer and used 2383 of these samples to evaluate their use for security applications.

Figure 2a shows an image of a nanostructured-silicon pattern. The image contains 1024 × 1024 pixels, has eight-bit resolution (256 levels) and was obtained by averaging eight frames acquired by CD-SEM. A variety of different morphologies were obtained, as shown in Fig. 2b. Figures 2b.i and 2b.iii show that the structural details in the patterns are as small as 9.23 nm.

Here, one minor remark is that the sizes and the layout of the original array of pillars have not been optimized so that the resultant security performances, described below, are maximized. Nevertheless, as shortly demonstrated, quite good properties have been obtained. This indicates that further advancements could be possible by engineering the original pillar (or not-like-a-pillar) structures to be collapsed, which could be an interesting future study. Meanwhile, we have experimentally confirmed that a proper dose of electron beam is necessary in order to induce versatile collapse of resist pillars; Figs. 2c and 2d show CD-SEM images when the dose was 30 and 40 µC/cm$^2$, respectively, indicating that too low or too high doses do not yield versatile resultant patterns.

To determine whether these patterns may be used as artifact metrics, we conducted the following analysis: A 512 pixel × 512 pixel, 8 bit (256 levels) greyscale image was extracted from the centre of an image of a pillar array and smoothed by an 11×11 median filter. In comparing any two patterns, $A(i,j)$ and $B(i,j)$, we first created a 'mask' pattern defined by

$$M(i,j) = \begin{cases} 1, & A(i,j) > T \text{ or } B(i,j) > T \\ 0 & \text{otherwise,} \end{cases} \qquad (1)$$

where *T* is a given threshold value. Because the patterns are fabricated by conventional lithographic processes, they consist of areas of varying heights. Therefore, two peaks appear in the statistics of pixel values, with a valley between the two peaks. Specifically, the number of times higher and lower peak values occur was approximately 130 and 80, respectively, and the incidence of the valley (i.e. the threshold *T*) between the two peaks was 90. Here, a remark is that the pixel value in images is given by 8 bit (0–255), and the particular values of 130, 80 and 90 are related to the greyscale pixel values of the given images. As indicated by Fig. 1c and Fig. 2, the greyscale value is related to the height of the nanostructured pattern. We do not calibrate the pixel value to the actual height (i.e. $A(i,j)$ and $B(i,j)$ are dimensionless values), but it does not cause any problem in this particular study. Also, 1 pixel occupies approximately a 3.3 nm square area.

By applying the mask, we obtain two images, $\hat{A}(i,j) = M(i,j) \times A(i,j)$ and $\hat{B}(i,j) = M(i,j) \times B(i,j)$, which means that we ignore regions where both patterns $A(i,j)$ and $B(i,j)$ are low (i.e. the pattern is not high). The correlation, or similarity, between the two patterns is evaluated by the Pearson correlation coefficient

$$R = \frac{\sum_i \sum_j \left[\hat{A}(i,j) - \overline{\hat{A}}\right]\left[\hat{B}(i,j) - \overline{\hat{B}}\right]}{\sqrt{\sum_i \sum_j \left[\hat{A}(i,j) - \overline{\hat{A}}\right]^2 \left[\hat{B}(i,j) - \overline{\hat{B}}\right]^2}}, \qquad (2)$$

where $\overline{\hat{A}}$ and $\overline{\hat{B}}$ indicate the average of $\hat{A}(i,j)$ and $\hat{B}(i,j)$, respectively. *R* is a dimensionless value. If *R* is negative, it is set to zero. Furthermore, each value of $A(i,j)$ and $B(i,j)$ is shifted between one and five pixels to the upper, lower, left and right side, and *R* is calculated for each shifted position. The maximum *R* from these positions (no shift, left, right, up and down shifts) is used as the similarity between $A(i, j)$ and $B(i, j)$.

We next calculate the false match rate (FMR) and false non-match rate (FNMR). FMR and FNMR are indicators of individuality and measurement stability, respectively. To calculate the FMR, all 2383 images were used. If the similarity given by Eq. (2) is greater than the given threshold, the two patterns are considered to be similar to each other, which is extremely likely to be a false decision. Since we had 2383 samples, we conducted 2383 × 2382 comparisons in calculating the FMR for threshold values between 0 and 1. To have an intuitive picture, suppose that only 1 case among the 2383 × 2382 comparisons resulted in a false decision. In this case, the 'error rate' would be $1/(2383 \times 2382) \approx 1.76 \times 10^{-7}$. From such calculation, the logarithmic scale for the FMR in the *y*-axis of Fig. 3 is naturally recognized. From Fig. 3, the FMR drops below the error rate of $10^{-6}$ as the threshold is just above zero, indicating that the occurrence of false decisions among all comparisons are extremely small.

To calculate the FNMR, we used images created from 100 images of each of the 74 samples. If the similarity is less than the given threshold, the two images are considered different from each other. In other words, identical samples are considered different, which is a false decision. The leftmost and rightmost curves in Fig. 3 show the FMR and FNMR, respectively. The FMR is substantially smaller than the FNMR, even with a smaller threshold value. In addition, the FMR and FNMR curves are well separated from each other, which means that it is possible to obtain sufficiently small FMR and FNMR by choosing adequate threshold values.

In addition, using the following method, we examined the clone match rate (CMR). Suppose that attackers capture the authentic device pattern and precisely fabricate a pattern in such a manner that an average over every $k \times k$ pixel area is essentially equivalent to the authentic device pattern. To quantify such a cloning process, we transformed each of the 2383 images into 'virtually cloned images'. Let the pixel value for a $k \times k$ area be denoted by $p(i, j)$. If the average value in this area is greater than the threshold $T$, we consider this area to have the higher average value (130). Thus, this pixel value of the virtually cloned image is $p'(i, j) = 130$. If the average value in the $k \times k$ area is less than or equal to the threshold $T$, we consider this area to have the lower average value (80). Thus, this pixel value of the virtually cloned image is $p'(i, j) = 80$. The clone image thus obtained is then compared with the authentic image using Eq. (2). If the similarity exceeds the threshold value, the clone successfully mimics the original. The CMR is calculated by performing the above evaluation for all the 2383 samples.

This scheme constitutes a very strict evaluation of the cloning resistance. Table 1 summarizes the assumed cloning technologies. For example, if $k = 3$, a $3 \times 3$ pixel area (or 'unit tile size') corresponds to a 10 nm × 10 nm square because, as mentioned earlier, 1 pixel occupies approximately a 3.3 nm square area, which may be regarded as state-of-the-art for the current nanofabrication technology. Nevertheless, the similarity cannot be greater than 0.4, as indicated by the calculated CMR shown by the purple dotted curve in Fig. 3. Considering that this value is considerably lower than the similarity between images of identical samples (i.e. the FNMR curve), such a clone does not pose a serious threat. In other words, the original authentic pattern is sufficiently random. Furthermore, were it possible to detect that a given pattern was formed from a combination of square units, it would presumably be determined that, based on such a feature, the pattern is a non-authentic device (or clone); this strategy is similar to liveness detection in biometrics[17]. Finally, note that the signal processing scheme described above is relatively simple yet requires highly skilled attackers. Based on these considerations and the results of the FMR, FNMR and CMR analysis, we conclude that the proposed nanostructured patterns based on the random collapse of resist could serve as superior nano-artifact metrics.

Finally, we put forward few remarks on the demonstrated principles and technologies.

First, the proposed principle is based on the 'uncertainty' inherent in conventional e-beam lithography technologies. Moreover, e-beam lithography is one of the widely spreading nanotechnologies including silicon fabrications. In this context, the proposed principle also has advantages in its general purpose properties or utilities.

Second is a comment regarding optical lithography. Optical lithography in silicon nanostructring is based on the so-called reticle, which is subjected to reduced-projection exposure. A reticle is four times larger in scale than the intended nanostructured pattern and is fabricated by e-beam lithography. It is impossible to fabricate patterns by optical lithography in the same resolution of the randomly collapsed silicon nanostructures demonstrated in this study. In other words, optical lithography may not be useful for attackers in copying the demonstrated devices.

The third remark concerns the term 'nano-artifact metrics'. In Ref. [1], Matsumoto *et al.* proposed 'artifact metrics'. The notion of artifact metrics is conceptually similar to 'biometrics', for which uniqueness in biological entities is utilized. Unlike biometrics, artifact metrics utilizes uniqueness in physical objects/things, physical processes or their combinations. We should emphasize that there are no implications such as 'defective patterns' in the word 'artifact'. One may imagine that 'nanoscale fingerprint' might be more appropriate than 'nano-artifact metrics'. However, the term 'fingerprint' implies information hiding, watermarking and their related technologies in the field of information security, which do not apply to our study. Furthermore, as discussed at the beginning of this paper, we showed four important requirements for artifact metrics to function (i.e. individuality, measurement stability, durability, clone resistance). Here, it should be noted that a total system as a whole is important, not just the elemental processes; this is another reason we describe the concept by 'nano-artifact metrics', which includes the notion of a total system while avoiding the use of 'nanoscale fingerprint'.

In summary, with the goal of exploiting the fundamental laws of physics to produce nano-artifact metrics, we demonstrated nano-artifact metrics based on the random collapse of resist pillars in e-beam lithography. As qualitative significance for information security, this study opens new design principles and degrees-of-freedom by exploiting uncertainty at the nanometre scale. Moreover, by developing sophisticated image preprocessing means and similarity indices for matching, the security performance of these metrics is further enhanced. This is also a qualitatively novel aspect for information security in the sense that the combinations of physical process and logical signal processing means provide new values. Note that our use of SEM technology is not a particularly important aspect of this study. To construct practical systems, many additional issues must be considered, such as reducing measurement cost and verifying interoperability for the case when different measurement devices are used for sample registration and authentication. Nevertheless, this study demonstrates that sufficiently good security performance can be achieved by the random collapse of resist in e-beam lithography; the enemy of silicon processing in previous studies turns out to be a strong enabler for information security in this study.

**Acknowledgements**

The authors thank Hitachi High-Technologies Cooperation in operating the CD-SEM. The authors thank Mr. Nakagawa from JOEL Co. Ltd. for his helpful discussions regarding the resolution limits of e-beam lithography. This work was supported in part by the Strategic Information and Communications R&D Promotion Programme (SCOPE) of the Ministry of Internal Affairs and Communications and by the Core-to-Core Program of the Japan Society for the Promotion of Science.

**Author contributions**

T. M., M. N. and M. O. directed the project; T. M., M. N., M. H. and N. T. designed the system architecture and experiments; M. H., Y. O. and M. I. performed experimental device fabrication; M. H. performed SEM measurements; T. M., K. H., R. S. and D. S. conducted image processing and performance analysis; and T. M. and M. N. wrote the paper.

**Competing financial interests**

The authors declare no competing financial interests.

**Table 1 | Assumptions for cloning technology.**

| Unit tile size of virtual clone image | Actual physical size |
|---|---|
| 15×15 pixels | 50-nm-square |
| 12×12 pixels | 40-nm-square |
| 9×9 pixels | 30-nm-square |
| 6×6 pixels | 20-nm-square |
| 3×3 pixels | 10-nm-square |

**Figure captions**

**Figure 1 | Nano-artifact metrics based on random collapse of resist in electron-beam lithography.** (a) Roadmap showing the minimum size of pillars formed by e-beam lithography. Using the phenomenon of randomly induced resist collapse, nano-artifact metrics contain length scales below the minimum dimension available in conventional lithography methods. (b) Schematic of array of pillars. (c) SEM image of collapsed resist.

**Figure 2 | Versatile morphology in silicon nanostructures obtained from collapsed resist.** (a) Example of entire region of fabricated silicon nanostructure. (b) Magnified view of several areas from the panel (a). The minimum feature size is indicated in each image. Note that feature sizes are smaller than the minimum feature size of the original array of pillars. In other words, the uncertainty obtained in this versatile morphology is less than that available directly by current technology. (c,d) Silicon nanostructure when the dose quantity in the e-beam lithography was (c) 30 and (d) 40 $\mu C/cm^2$, respectively. Too low or too high doses do not yield versatile resultant patterns.

**Figure 3 | Evaluation of security performance.** Error rate as a function of the threshold. False match rate (FMR) and false non-match rate (FNMR) are labelled. The curves labelled 60, 50, 40, 30, 20 and 10 nm are the clone match rate (CMR) for the given minimum unit tile size.

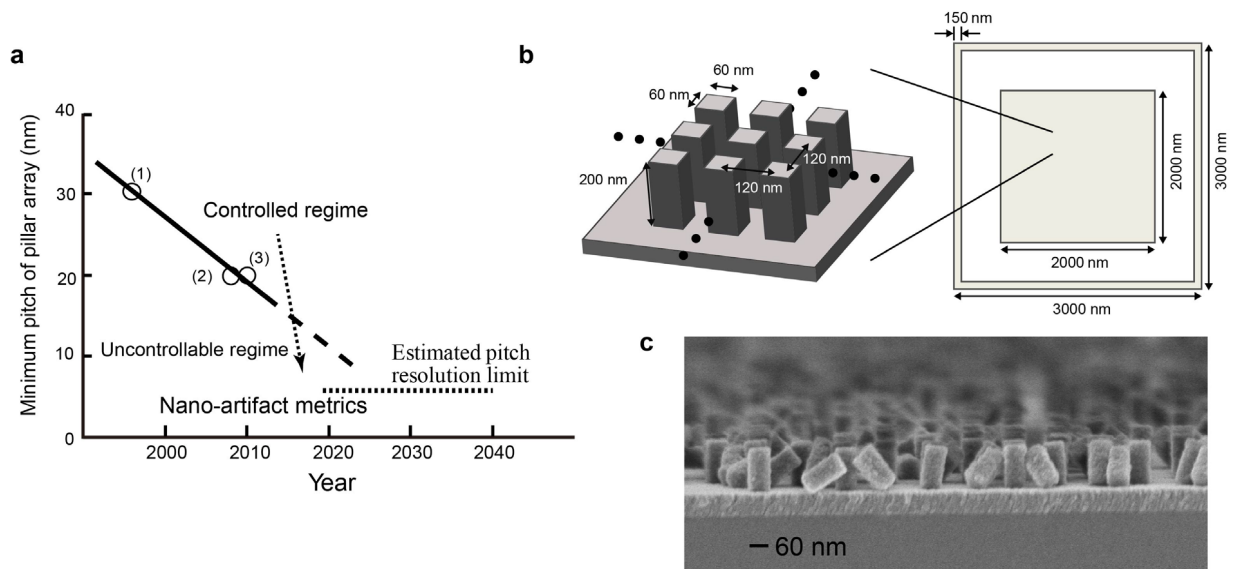

Figure 1

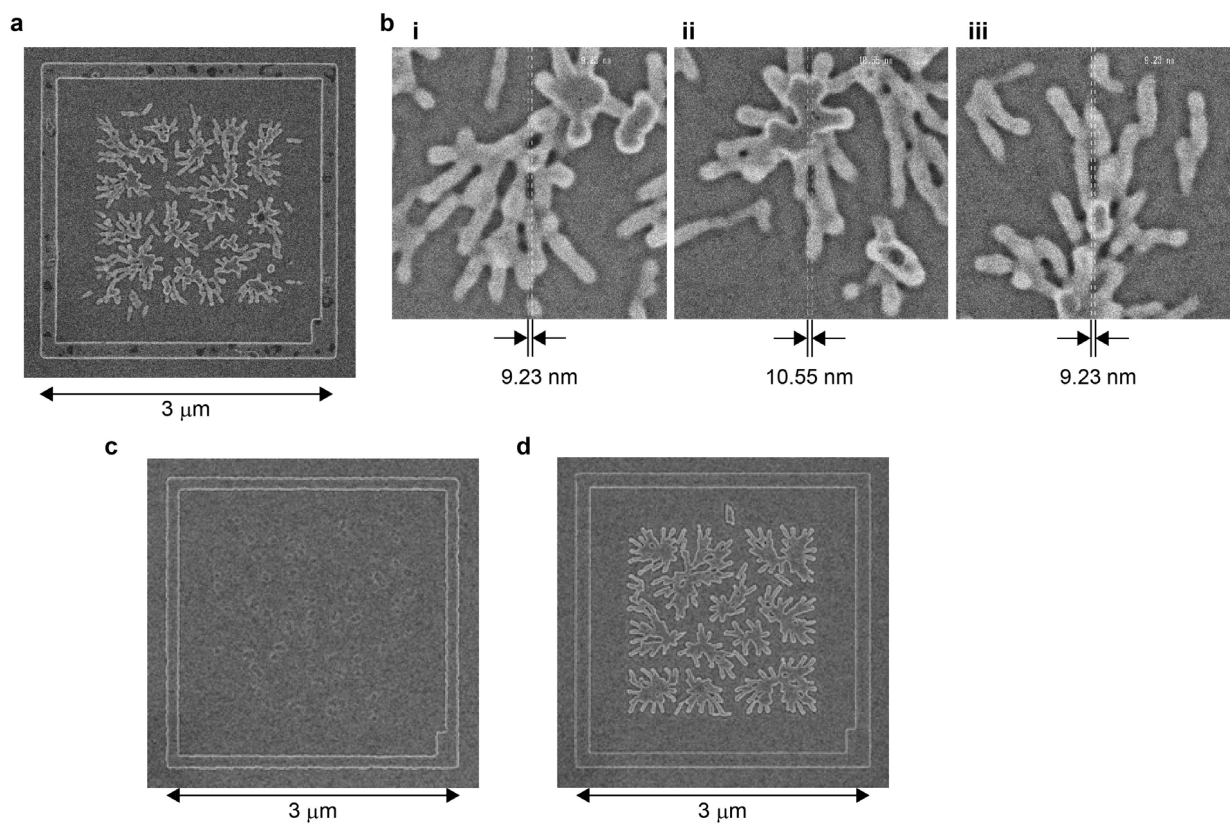

Figure 2

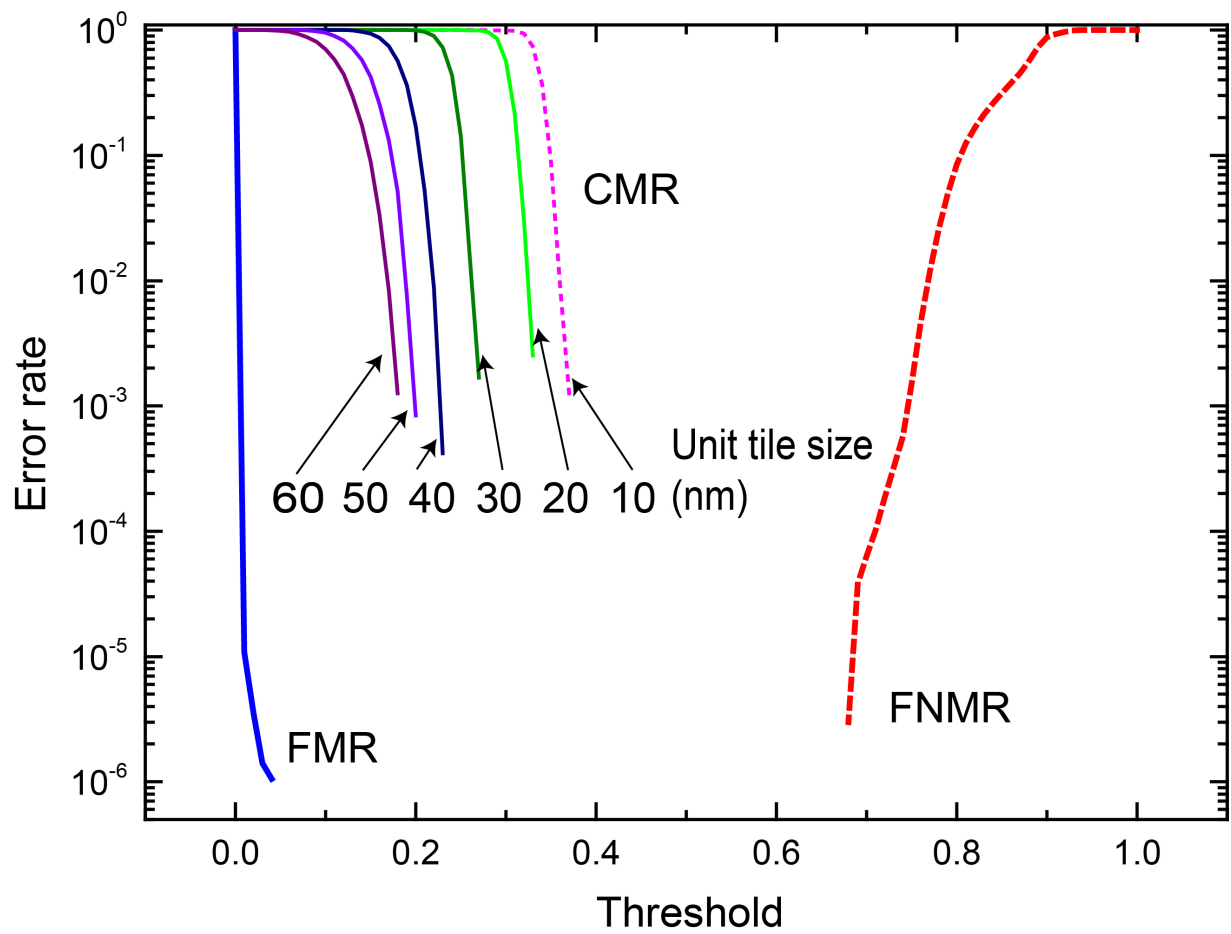

Figure 3